\documentclass[11pt]{article}

\makeatletter
\renewcommand*\@fnsymbol[1]{\the#1}
\makeatother

\usepackage{amsmath}
\usepackage{amssymb}
\usepackage{amsfonts}
\usepackage{amsthm}
\usepackage[english]{babel}
\usepackage{latexsym}
\usepackage[hmargin=1.8cm,vmargin=2.4cm]{geometry}
\usepackage{enumerate}
\usepackage{nicefrac}
\usepackage{mathrsfs}
\usepackage[affil-it]{authblk}
\usepackage{datetime}
\usepackage{comment}
\usepackage{color}
\usepackage{graphicx}
\usepackage{epstopdf}

\theoremstyle{plain}
\newtheorem{theorem}{Theorem}[section]

\theoremstyle{definition}

\newcommand{\Interior}{\mathop{\rm int}\nolimits}

\newcommand{\bd}{\mathop{\rm bd}\nolimits}

\newcommand{\risk}{\rho}
\newcommand{\Risk}{{\mathcal E}}

\newcommand{\cA}{\mathcal A}
\newcommand{\cK}{\mathcal K}
\newcommand{\cL}{\mathcal L}
\newcommand{\cM}{\mathcal M}
\newcommand{\cN}{\mathcal N}
\newcommand{\cS}{\mathcal S}
\newcommand{\cX}{\mathcal X}

\newcommand{\cB}{\mathcal B}

\newcommand{\cU}{\mathcal U}

\newcommand{\R}{\mathbb R}
\newcommand{\N}{\mathbb N}

\begin{document}

\title{
A continuous selection for optimal portfolios\\
under convex risk measures does not always exist
}

\author{
\sc{Michel Baes}
}
\affil{Department of Mathematics\\
ETH Zurich, Switzerland\\
mbaes@math.ethz.ch}
\author{
\sc{Cosimo Munari}
}
\affil{Center for Finance and Insurance and Swiss Finance Institute\\
University of Zurich, Switzerland\\
cosimo.munari@bf.uzh.ch}

\date{\usdate\today}

\maketitle

\abstract{
One of the crucial problems in mathematical finance is to mitigate the risk of a financial position by setting up hedging positions of eligible financial securities. This leads to focusing on set-valued maps associating to any financial position the set of those eligible payoffs that reduce the risk of the position to a target acceptable level at the lowest possible cost. Among other properties of such maps, the ability to ensure lower semicontinuity and continuous selections is key from an operational perspective. It is known that lower semicontinuity generally fails in an infinite-dimensional setting. In this note we show that neither lower semicontinuity nor, more surprisingly, the existence of continuous selections can be a priori guaranteed even in a finite-dimensional setting. In particular, this failure is possible under arbitrage-free markets and convex risk measures.

}

\bigskip

\noindent \textbf{Keywords}: risk measures, portfolio selection, perturbation analysis, continuous selections.

\medskip

\noindent {\bf Mathematics Subject Classification}: 91B30, 91B32



\parindent 0em \noindent


\section{Introduction}

This note deals with the existence of continuous selections for a class of optimal set mappings that play an important role in several areas of mathematical finance, including capital adequacy, pricing, hedging, and capital allocation. In that context, one if often confronted with the problem of mitigating the risk of a given financial position by setting up a suitable capital buffer whose function is to absorb future larger-than-expected losses. This capital reserve is usually held in the form of a portfolio of some financial securities called the eligible assets. The optimal set mappings under consideration associate to each financial position precisely the set of all payoffs of eligible assets that allow to confine risk within an acceptable level of security at the lowest possible cost.

\medskip

A thorough analysis of qualitative robustness for such set-valued maps has been recently provided in~\cite{BaesKochMunari2017}. Among the various stability properties, lower semicontinuity proves to be of cardinal importance in that it ensures that any optimal payoff of eligible assets remains close to being optimal after a slight misestimation of misspecification of the underlying financial position. However, one of the key findings of~\cite{BaesKochMunari2017} was that lower semicontinuity is typically not satisfied. The counterexamples to lower semicontinuity provided there exploited in an essential way the infinite-dimensional structure of the underlying model space. At the same time, the lower semicontinuity results established there suggest that, under appropriate convexity assumptions, lower semicontinuity might not be too difficult to ensure in a finite-dimensional setting. It is therefore natural to ask whether, by restricting the attention to finite-dimensional model spaces and by working in a suitable convex environment, one may always guarantee lower semicontinuity.

\medskip

From an operational perspective, the existence of continuous selections constitutes another key property of the above optimal set mappings that allows to associate to each financial position a unique portfolio of eligible assets in such a way that a slight perturbation of the underlying financial position does not engender a dramatic change in the structure of the corresponding optimal portfolio. Recall that, for convex-valued maps, the existence of continuous selections is automatically implied by lower semicontinuity by virtue of Michael's Theorem (\cite[Theorem~17.66]{AliprantisBorder2006}). Hence, one may hope to always have continuous selections in a finite-dimensional setting, at least under convexity, even in the case that lower semicontinuity failed. This question was not addressed in~\cite{BaesKochMunari2017}.

\medskip

In this note we provide concrete examples of optimal set mappings of the above type in a finite-dimensional model space that (1) fail to be lower semicontinuous but admit a continuous selection (2) fail to admit a continuous selection. Besides their intrinsic mathematical interest, our examples raise a serious warning in the above-mentioned fields of application: a cost or risk minimization problem under a convex risk measure in a finite-dimensional space need not allow for a robust way to select optimal portfolios of eligible assets. Hence, a case-by-case analysis is required to establish whether a special choice of a (convex) risk measure leads to robust optimal selections or not.

\medskip

This note is structured as follows. In Section~2 we introduce our mathematical setting and define the relevant class of optimal set mappings. In Section~3 we show that an optimal set mapping in a finite-dimensional setting may fail to be lower semicontinuous but still admit continuous selections. Section~4, which builds on the example discussed in the previous section, establishes that an optimal set mapping in a finite-dimensional setting may even fail to admit continuous selections.


\section{The optimal set mapping}

We introduce our optimal set mapping by adopting the same notation of~\cite{BaesKochMunari2017}, to which we refer for more details about the non-mathematical aspects of our problem.

\medskip

Consider a one-period economy with dates $t=0$ (the initial date) and $t=1$ (the terminal date). {\em Financial positions} at the terminal date are represented by the elements of a (real) Hausdorff topological vector space $\cX$, which we assume to be partially ordered by a convex cone $\cX_+$. Within the space of position one identifies a set $\cA$ of {\em acceptable} (from the point of view of financial regulators) or {\em desirable} (from the point of view of risk or portfolio managers) positions. We denote by $\cM$ a finite-dimensional linear subspace of $\cX$, whose elements represent the {\em payoffs} of a finite number of financial assets that are used to push unacceptable positions into the target set $\cA$. The space $\cM$ is endowed with the relative topology induced by $\cX$. Each payoff in $\cM$ carries a certain {\em price} that is represented by a linear functional $\pi:\cM\to\R$.

\medskip

From a capital management perspective it is important to know at which cost a certain financial position can be made acceptable. This leads to studying the {\em optimal value function} $\risk:\cX\to\overline{\R}$ defined by
\[
\risk(x) := \inf\{\pi(z) \,; \ z\in\cM, \ z+x\in\cA\}.
\]
In the financial literature the above function is usually referred to as a {\em risk measure}. The interested reader can consult~\cite{ArtznerDelbaenEberHeath1999}, \cite{ArtznerDelbaenKoch2009}, \cite{FarkasKochMunari2014}, \cite{FarkasKochMunari2015}, \cite{FoellmerSchied2011} for a variety of results on risk measures and discussions on their financial relevance in different areas of mathematical finance.

\medskip

The {\em optimal set mapping} associated to the above parametric optimization problem is the set-valued map $\Risk:\cX\rightrightarrows\cM$ given by
\[
\Risk(x) := \{z\in\cM \,; \ z+x\in\cA, \ \pi(z)=\risk(x)\}.
\]
Any element of $\Risk(x)$ is referred to an an {\em optimal payoff} for $x$. We refer to~\cite{BaesKochMunari2017} for a comprehensive study of qualitative robustness for the above mappings. As mentioned in the introduction, one of the main findings of that paper was to show that, in many relevant cases, the map $\Risk$ fails to be lower semicontinuous. This is true even if the acceptance set $\cA$ is assumed to be convex. Recall that $\Risk$ is {\em lower semicontinuous} at some $x\in\cX$ whenever for any open set $\cU\subset\cM$ satisfying $\Risk(x)\cap\cU\neq\emptyset$ there exists a neighborhood $\cU_x\subset\cX$ of $x$ such that
\[
y\in\cU_x \ \implies \ \Risk(y)\cap\cU\neq\emptyset.
\]
Intuitively speaking, this means that any optimal payoff for $x$ remains close to being optimal after a slight perturbation of $x$.

\medskip

However, all the counterexamples to lower semicontinuity exhibited in~\cite{BaesKochMunari2017} use in a critical way the infinite dimensionality of the underlying ambient space. In fact, none of them can be reproduced in a finite-dimensional setting due to the general results from the same paper. More precisely, the acceptance sets used in the counterexamples become polyhedral sets once restricted to finite dimension and lower semicontinuity is always ensured under polyhedrality by virtue of~\cite[Theorem~5.12]{BaesKochMunari2017}. It remained thus open whether, especially for convex acceptance sets, one can still find counterexamples to lower semicontinuity and, more generally, to the existence of continuous selections for $\Risk$ in a finite-dimensional setting.

\medskip

We aim to enrich the results of~\cite{BaesKochMunari2017} by showing that not only might lower semicontinuity fail for a convex acceptance set in a finite-dimensional model space, but we might even fail to find continuous selections for the optimal set mapping. This is also true if we impose the following requirements:
\begin{enumerate}
  \item[(R1)] $\cA$ is closed, convex, contains zero and satisfies the monotonicity property
\[
x\in\cA, \ y\in x+\cX_+ \ \implies \ y\in\cA.
\]
  \item[(R2)] $\cM$ admits no arbitrage opportunity, i.e.
\[
z\in\cM\cap\cX_+\setminus\{0\} \ \implies \ \pi(z)>0.
\]
  \item[(R3)] $\risk$ is finite and continuous.
\end{enumerate}
In this case, we will say that $(\cA,\cM,\pi)$ is {\em admissible}. The assumptions on $\cA$ are standard in the risk measure literature. In particular, by stipulating that any aggregation of acceptable positions remains acceptable, the property of convexity is often viewed to provide the mathematical translation of the economics principle of diversification, according to which aggregation should always improve security. The monotonicity property requires that any position dominating some acceptable position should also be deemed acceptable. The absence of arbitrage opportunities, which corresponds to the strict positivity of the pricing functional $\pi$, is universally encountered in the mathematical finance literature. Finally, assuming that $\risk$ be finite and continuous makes the question of lower semicontinuity meaningful and our search for a counterexample more challenging.


\section{Convexity does not ensure lower semicontinuity}

Throughout this section we assume that $\cX=\R^3$. Our aim is to construct an admissible triple $(\cA,\cM,\pi)$ and exhibit a vector $x\in\R^3$ such that $\Risk$ fails to be lower semicontinuous at $x$. We split our construction in several steps.

\subsubsection*{The basic set}

\begin{figure}[th]
\centering
\includegraphics[width=0.5\textwidth]{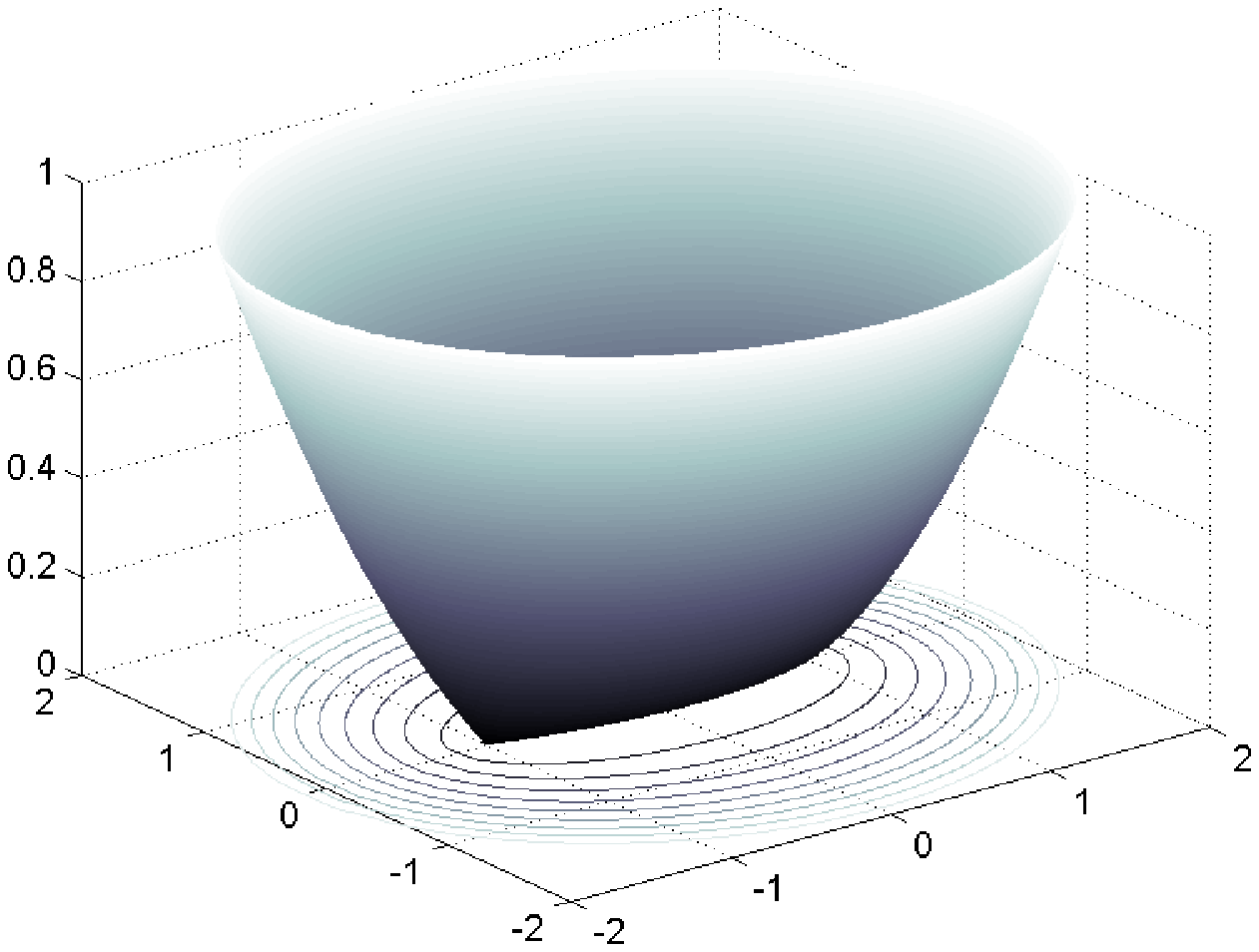}
\caption{The set $\cB_r$ for $r=2$.}
\label{fig:Boat}
\end{figure}

The acceptance set will be obtained by a suitable extension and rotation applied to the following basic set. Fix $r>0$ and define a subset of $\R^3$ by setting
\[
\cB_r = \bigg\{x\in\R^2\times(0,1] \,; \ \frac{x_1^2}{1+r^2x_3}+\frac{x_2^2}{r^2x_3}\leq 1\bigg\}\cup
\{x\in\R^3 \,; \ x_1\in[-1,1], \ x_2=x_3=0\}.
\]
The lower boundary of this set has the boat-like shape depicted in Figure~\ref{fig:Boat} with the projection of some level sets. For every $h\in[0,1]$ the slice $\cS(h)=\{x\in\cB_r \,; \ x_3=h\}$ is an ellipsoid centered in $0$ with axes parallel to the canonical vectors $e_1$ and $e_2$ and of lengths $2\sqrt{1+r^2h}$ and $2r\sqrt{h}$, respectively. In particular, the slice $\cS(0)$ is a degenerated ellipsoid whereas the slice $\cS(1)$ contains a circle of radius $r$.

The set $\cB_r$ is clearly closed and is easily seen to be convex. Indeed, since the function $g:\R\times(0,\infty)\to\R$ defined by $g(s,t)=s^2/t$ is convex, it follows that $\lambda x+(1-\lambda)y\in\cB_r$ for every $x,y\in\cB_r$ with $x_3>0$ and $y_3>0$ and $\lambda\in[0,1]$. The same conclusion holds for every $x,y\in\cB_r$ by closedness.

\medskip

For every given radius $0<R\leq\tfrac{r^2}{2\sqrt{1+r^2}}$ consider the ice-cream cone
\[
\cK_R = \{x\in\R^3 \,; \ x_1^2+x_2^2\leq R^2x_3^2\}.
\]
We claim that $\cB_r$ satisfies
\begin{equation}
\label{eq: monotonicity boat 0}
(\cB_r+\cK_R)\cap\{x\in\R^3 \,; \ x_3\leq1\} = \cB_r.
\end{equation}

\smallskip

To prove this, consider the convex function $f_r:\R^2\to\R$ given by
\begin{equation}
\label{eq: auxiliary function boundary boat}
f_r(x_1,x_2) = \frac{x_1^2+x^2_2-1+\sqrt{(x_1^2+x^2_2-1)^2+4x^2_2}}{2r^2}.
\end{equation}

After some elementary manipulations, one can show that $\cB_r$ is nothing but a section of the epigraph of $f_r$, namely
\[
\cB_r = \{x\in\R^2\times[0,1] \,; \ f_r(x_1,x_2)\leq x_3\}.
\]

\smallskip

For every $x\in\bd\cB_r$ with $x_3\in(0,1)$ one can easily verify that
\[
\partial_1 f_r(x_1,x_2)=
\frac{1}{2r^2}\frac{2x_1f_r(x_1,x_2)}{f_r(x_1,x_2)-(x_1^2+x^2_2-1)} \ \ \ \mbox{and} \ \ \
\partial_2 f_r(x_1,x_2)=
\frac{1}{2r^2}\frac{2x_2(f_r(x_1,x_2)+2)}{f_r(x_1,x_2)-(x_1^2+x^2_2-1)}.
\]
Since $f_r(x_1,x_2)=x_3$ and $x_1^2+x^2_2-1=r^2x_3-x_2^2/(r^2x_3)$, we infer that
\begin{eqnarray*}
\|\nabla f_r(x_1,x_2)\|_2^2
&=&
\frac{1}{4r^4}\frac{4(x_1^2+x_2^2)f_r^2(x_1,x_2)+16x_2^2f_r(x_1,x_2)+16x_2^2}
{(f_r(x_1,x_2)-(x_1^2+x^2_2-1))^2} \\
&=&
\frac{1}{4r^4}\frac{16r^4x_3^2(1+r^2x_3)}{r^4x_3^2+x_2^2} \\
&\leq&
\frac{4(1+r^2)}{r^4}.
\end{eqnarray*}

\smallskip

It follows that~\eqref{eq: monotonicity boat 0} is satisfied. To see this, take any $x\in\cB_r$ with $x_3\in(0,1)$ and $y\in\cK_R$ such that $x_3+y_3\leq1$. The uniform bound on the norm of the gradient established above allows us to write
\[
f_r(x_1+y_1,x_2+y_2)
\leq
f_r(x_1,x_2)+\frac{2\sqrt{1+r^2}}{r^2}\sqrt{y_1^2+y_2^2}
\leq
x_3+\frac{2\sqrt{1+r^2}}{r^2}R y_3
\leq
x_3+y_3
\]
where we used that $R\leq\tfrac{r^2}{2\sqrt{1+r^2}}$. By continuity, the inequality $f_r(x_1+y_1,x_2+y_2)\leq x_3+y_3$ can be extended to any $x\in\cB_r$ with $x_3=0$ and $y\in\cK_R$ such that $y_3\leq1$. This establishes~\eqref{eq: monotonicity boat 0}.

\subsubsection*{The rotation}

Consider the isometry $\Phi:\R^3\to\R^3$ defined by
\begin{equation}
\label{eq: rotation}
\Phi(x)=
\begin{pmatrix}
\tfrac{1}{\sqrt{2}} & \tfrac{1}{\sqrt{6}} & \tfrac{1}{\sqrt{3}} \\
-\tfrac{1}{\sqrt{2}} & \tfrac{1}{\sqrt{6}} & \tfrac{1}{\sqrt{3}} \\
0 & -\tfrac{\sqrt{2}}{\sqrt{3}} & \tfrac{1}{\sqrt{3}} \\
\end{pmatrix}
\begin{pmatrix}
x_1 \\
x_2 \\
x_3
\end{pmatrix}.
\end{equation}
It can be easily verified that $\Phi$ is the composition of a clockwise rotation of an angle $\vartheta=\pi/4$ around the unit vector $e_3$ with a clockwise rotation of an angle $\vartheta$ such that $\sin(\vartheta)=\sqrt{2}/\sqrt{3}$ and $\cos(\vartheta)=1/\sqrt{3}$ around the unit vector $(1/\sqrt{2},-1/\sqrt{2},0)$.

\subsubsection*{The acceptance set}

Let $r=3$ and consider the set $\cA\subset\R^3$ defined by
\[
\cA = \Phi(\cB_r)+\R^3_+.
\]
Recall that $\cB_r$ is a compact and convex set containing zero. As a result, we immediately see that $\cA$ is closed, convex, and contains zero. Moreover, $\cA$ satisfies the monotonicity property by definition. Hence, requirement (R1) is fulfilled.

\subsubsection*{The payoff space}

The space of payoffs is the linear subspace of $\R^3$ given by
\[
\cM = \Phi(\cN)
\]
where $\cN=\{w\in\R^3 \,; \ w_2=0\}$. It is immediate to see that $\cM$ is spanned by $(1,1,1)=\Phi(0,0,\sqrt{3})$ and $(1,-1,0)=\Phi(\sqrt{2},0,0)$. In addition, the pricing functional is the linear functional $\pi:\cM\to\R$ given by
\[
\pi(z) = z_3.
\]
In particular, we have $\pi(1,1,1)=1$ and $\pi(1,-1,0)=0$. To show that $\cM$ admits no arbitrage, take any nonzero $z\in\cM\cap\R^3_+$ and note that
\[
z = \Phi(w) = \left(\frac{w_1}{\sqrt{2}}+\frac{w_3}{\sqrt{3}},
-\frac{w_1}{\sqrt{2}}+\frac{w_3}{\sqrt{3}},\frac{w_3}{\sqrt{3}}\right)
\]
for a suitable $w\in\cN$. Since $z$ is positive but nonzero, we must have $w_3>0$ and therefore $\pi(z)=z_3>0$. This shows that (R2) holds.

\medskip

To show finiteness and continuity of $\risk$, note first that
\begin{eqnarray*}
\risk(x)
&=&
\inf\{\pi(z) \,; \ z\in\cM, \ z+x\in\Phi(\cB_r)+\R^3_+\} \\
&=&
\inf\{\pi(\Phi(w)) \,; \ w\in\cN, \ w+\Phi^{-1}(x)\in\cB_r+\Phi^{-1}(\R^3_+)\} \\
&=&
\tfrac{1}{\sqrt{3}}\inf\{w_3 \,; \ w\in\cN, \ w+\Phi^{-1}(x)\in\cB_r+\Phi^{-1}(\R^3_+)\}.
\end{eqnarray*}

\smallskip

Set $R=\sqrt{2}$ so that $\Phi^{-1}(\R^3_+)\subset\cK_R$. Since $R\leq\tfrac{r^2}{2\sqrt{1+r^2}}$, it follows from~\eqref{eq: monotonicity boat 0} that
\[
(\cB_r+\Phi^{-1}(\R^3_+))\cap\{x\in\R^3 \,; \ x_3\leq1\} = \cB_r\,.
\]
As a result, we see that $\risk(0)=0$. Now, take any $x\in\R^3$ and note that
\[
\|x\|_\infty e \geq x \geq -\|x\|_\infty e,
\]
where $e=(1,1,1)$. Then, it is not difficult to verify that
\[
-\|x\|_\infty = \risk(0)-\pi(\|x\|_\infty e) = \risk(\|x\|_\infty e) \leq \risk(x) \leq \risk(-\|x\|_\infty e) = \risk(0)+\pi(\|x\|_\infty e) = \|x\|_\infty.
\]
This shows that $\risk$ is finitely valued and (globally Lipschitz) continuous. Hence, requirement (R3) is also satisfied and the triple $(\cA,\cM,\pi)$ is admissible.

\subsubsection*{The failure of lower semicontinuity}

First of all, note that for every $x\in\R^3$ we can write
\begin{eqnarray*}
\Risk(x)
&=&
\{z\in\cM \,; \ z+x\in\cA, \ \pi(z)=\risk(x)\} \\
&=&
\{\Phi(w) \,; \ w\in\cN, \ w+\Phi^{-1}(x)\in\cB_r+\Phi^{-1}(\R^3_+), \ \tfrac{w_3}{\sqrt{3}}=\risk(x)\} \\
&=&
\Phi(\{w\in\cN \,; \ w+\Phi^{-1}(x)\in\cB_r+\Phi^{-1}(\R^3_+), \ \tfrac{w_3}{\sqrt{3}}=\risk(x)\}).
\end{eqnarray*}

\smallskip

Consider now the sequence in $\cB_r$ with general term
\[
y^{(n)} = \left(0,\frac{r}{\sqrt{n}},\frac{1}{n}\right).
\]
Note that $\Phi(y^{(n)})\to0$ and for every $n\in\N$ we have $\risk(\Phi(y^{(n)}))=0$ and
\[
\Risk(\Phi(y^{(n)})) = \Phi(\{0\}) = \{0\}.
\]
At the same time, recalling that $\risk(0)=0$, it clearly holds
\begin{eqnarray*}
\Risk(0)
&=&
\Phi(\{w\in\R^3 \,; \ w_1\in[-1,1], \ w_2=w_3=0\}) \\
&=&
\{\lambda\Phi(-1,0,0)+(1-\lambda)\Phi(1,0,0) \,; \ \lambda\in[0,1]\}\,.
\end{eqnarray*}
This shows that $\Risk$ fails to be lower semicontinuous at $0$. In fact, a slight perturbation of $0$ may cause the corresponding set of optimal payoffs to shrink from an infinite set to just a singleton!

\section{Convexity does not ensure continuous selections}

Albeit not lower semicontinuous, the optimal set mapping in the preceding example is easily seen to admit a continuous selection. In this section we show that, by properly modifying the above acceptance set, we may be unable to ensure the existence of continuous selections as well. This is also possible under the constraint that $(\cA,\cM,\pi)$ be an admissible triple.

\medskip

To this effect, we follow the general construction of the preceding example. The only difference is that we will ``twist'' the set $\cB_r$ in a suitable way. This modification will require some technical preliminary work to ensure convexity, which was automatic in the above example. In the sequel, for any $r>0$ and any subset $E\subset\R^2$ we will consider sets of the form
\[
\cB_r(E) = \{(u\sqrt{1+r^2x_3},vr\sqrt{x_3},x_3)\in\R^3 \,; \ (u,v)\in E, \ x_3\in[0,1]\}.
\]

\subsubsection*{The auxiliary set $\cB_r(E)$}

Fix $r>0$. In this section we focus on sets $\cB_r(E)$ where
\[
E = \{(u,v)\in\R^2 \,; \ \alpha(u-a)^2+\beta v^2\leq1\}
\]
for given $a\in\R$ and $\alpha,\beta>0$. Note that we can always write
\[
\cB_r(E) = \{x\in\R^3 \,; \ F(x)\leq0, \ x_3\in[0,1]\}
\]
where $F:\R^3\to\R$ is specified by setting
\begin{equation}
\label{eq: representation piece}
F(x) = \frac{r^2x_3}{\beta}(\alpha(u(x_1,x_3)-a)^2-1)+x_2^2
\end{equation}
with $u(x_1,x_3)=\tfrac{x_1}{\sqrt{1+r^2x_3}}$.

\medskip

Now, assume that $|a|<\tfrac{1}{\sqrt{\alpha}}$ so that $0\in\Interior E$. In this case, $F$ is infinitely differentiable and satisfies
\begin{equation}
\label{eq: convexity tilted boat}
\partial_3 F(x) = \frac{r^2}{\beta} \left((\alpha(u(x_1,x_3)-a)^2-1)-\frac{\alpha r^2 x_3(u(x_1,x_3)-a)u(x_1,x_3)}{1+r^2x_3} \right) \neq 0
\end{equation}
whenever $x\in\cB_r(E)$ with $x_3>0$. Indeed, if we found $\partial_3 F(x)=0$ for some $x\in\cB_r(E)$ with $x_3>0$ we would produce the contradiction
\begin{eqnarray*}
0
&\geq&
\alpha(u(x_1,x_3)-a)^2-1 \\
&=&
r^2x_3(1-\alpha a^2)+\alpha r^2x_3u(x_1,x_3)a \\
&\geq&
r^2x_3(1-\alpha|a(a-u(x_1,x_3))|) \\
&\geq&
r^2x_3(1-\sqrt{\alpha}|a|) \\
&>&
0
\end{eqnarray*}

\smallskip

where we used that $|a-u(x_1,x_3)|\leq\tfrac{1}{\sqrt{\alpha}}$ in the second-to-last inequality and $|a|<\tfrac{1}{\sqrt{\alpha}}$ in the last inequality. As a result, \eqref{eq: convexity tilted boat} holds. Then, by the Implicit Function Theorem, there exists an open set $\cU\subset\R^2$ and a continuously differentiable function $f:\cU\to(0,\infty)$ for which
\[
\cB_r(E)\cap\{x\in\R^3 \,; \ x_3>0\} = \{x\in\R^3 \,; \ f(x_1,x_2)\leq x_3, \ x_3\in(0,1]\}.
\]
Of course, it is not difficult to see that we can extend this function continuously to obtain
\begin{equation}
\label{eq: implicit function}
\cB_r(E) = \{x\in\R^3 \,; \ f(x_1,x_2)\leq x_3, \ x_3\in[0,1]\}.
\end{equation}

\smallskip

{\bf Convexity of $\cB_r(E)$}. We claim that $\cB_r(E)$ is convex whenever $|a|<\tfrac{1}{\sqrt{\alpha}}$. To show this, we shall use the following result by~\cite{Crouzeix1980}.

\begin{theorem}
\label{theo: crouzeix}
Let $f:\R^n\to\R\cup\{+\infty\}$ be a lower semicontinuous quasiconvex function. Then, $f$ is convex if and only if the function $\sigma_s:\R\to\R\cup\{\infty\}$ defined by
\[
\sigma_s(t) = \sup\left\{\sum_{i=1}^n s_ix_i \,; \ x\in\R^n, \ f(x)\leq t\right\}
\]
is concave for every $s\in\R^n$.
\end{theorem}

\smallskip

Note first that $f$ is lower semicontinuous and quasiconvex by construction. We use Theorem~\ref{theo: crouzeix} to prove its convexity. For every $t\in[0,1]$ the sublevel set $\cL(t)=\{(x_1,x_2)\in\R^2 \,; \ f(x_1,x_2)\leq t\}$ is easily seen to satisfy
\[
\cL(t) = \{(u\sqrt{1+r^2t},vr\sqrt{t}) \,; \ (u,v)\in\R^2, \ \alpha(u-a)^2+\beta v^2\leq 1\}.
\]
This set is an ellipsoid and its support function can be computed explicitly. For $s\in\R^2$ we indeed have
\[
\sigma_s(t) = \sup_{(x_1,x_2)\in\cL(t)}s_1x_1+s_2x_2 = \sqrt{\frac{s_1^2(1+r^2t)}{\alpha}+\frac{s_2^2r^2t}{\beta}}+as_1\sqrt{1+r^2t}.
\]
As a function of $t$, we have a sum of two square roots of affine functions. If $as_1\geq0$, these two terms are concave. Assume, then, that $as_1<0$. The second derivative of the first term is
\[
-\frac{1}{4}\left(\frac{s_1^2r^2}{\alpha}+\frac{s_2^2r^2}{\beta}\right)^2\left(\frac{s_1^2(1+r^2t)}{\alpha} + \frac{s_2^2r^2t}{\beta}\right)^{-3/2},
\]
which can be seen to become more and more negative as $s_2^2$ increases. So, to establish the concavity of $\sigma_s$, it suffices to consider $s_2=0$. In this case, we get
\[
\sigma_s(t) = \sqrt{\frac{s_1^2(1+r^2t)}{\alpha}} - \sqrt{a^2s^2_1(1+r^2t)} =  |s_1|\sqrt{1+r^2t}\left(\frac{1}{\sqrt{\alpha}}-a \right),
\]
which is concave as $a<\tfrac{1}{\sqrt{\alpha}}$. This proves that $f$ is convex and, hence, that $\cB_r(E)$ is also convex.

\medskip

The convexity of $\cB_r(E)$ holds even if $a\sqrt{\alpha}=1$. To see this, define
\[
E(n) = \{(u,v)\in\R^2 \,; \ \alpha\left(u-a_n\right)^2+\beta v^2\leq 1\}
\]
for any $n\in\N$, where $a_n\uparrow a$ under the assumption that none of the ellipsoids $E(n)$ is empty. Now, let $x\in\cB_r(E)$. Then, there exists a sequence $x^{(n)}\to x$ such that $x^{(n)}\in\cB_r(E(n))$ for every $n\in\N$. This result is trivial when $x_3=0$. Otherwise, setting $v(x_2,x_3)=x_2/(r\sqrt{x_3})$ so that $\alpha(u(x_1,x_3)-a)^2+\beta v(x_2,x_3)^2\leq 1$, we simply take
\[
u_n(x_1,x_3)=u(x_1,x_3)-a+a_n, \ \ \ v_n(x_2,x_3)=v(x_2,x_3), \ \ \ x^{(n)}_3=x_3
\]
and construct the corresponding point $x^{(n)}$ of $E(n)$. Conversely, if $x^{(n)}\in\cB_r(E(n))$ defines a sequence converging to $x$ with $x_3>0$, then $x\in\cB_r(E)$ by continuity of the functions $u$ and $v$. This property is also immediately verified when $x_3=0$. As a result, if we let $a_n\uparrow a$, then we see that $\cB_r(E)$ is convex due to the convexity of each of the sets $\cB_r(E(n))$.

\medskip

{\bf Curvature of $\cB_r(E)$}. Finally, for our later construction, we need to provide some estimates on the norm of the gradient of $f$. For any $x\in\R^3$ with $x_3>0$ it follows from the Implicit Function Theorem that
\[
||\nabla f(x_1,x_2)||_2 = \frac{\sqrt{\partial_1 F(x)^2+\partial_2 F(x)^2}}{\left|\partial_3 F(x)\right|}.
\]
Now, assume that $a\geq0$ and $\alpha<5$ and take $r>\max\{\sqrt{2},\tfrac{1}{\sqrt{5-\alpha}}\}$. We claim that
\begin{equation}
\label{eq: bound gradient tilted 1}
\max\left\{||\nabla f(x_1,x_2)||_2 \,; \ x\in\bd\cB_r(E), \ 0\leq u(x_1,x_3)-a\leq\frac{1}{\sqrt{\alpha}}\right\} \leq \frac{2}{r}
\end{equation}
and, under the assumption $8>9\alpha a^2$,
\begin{equation}
\label{eq: bound gradient tilted 2}
\max\left\{||\nabla f(x_1,x_2)||_2 \,; \ x\in\bd\cB_r(E), \ -\frac{1}{\sqrt{\alpha}}\leq u(x_1,x_3)-a\leq\frac{1}{\sqrt{\alpha}}\right\} \leq \frac{16\max\{\tfrac{\alpha r^2}{r^2+1},1\}}{r(8-9\alpha a^2)}.
\end{equation}
Note that we can restrict the above optimization domains to those $x\in\bd\cB_r(E)$ such that $f(x_1,x_2)=x_3=1$ by convexity. After a few elementary rearrangements, we get
\begin{eqnarray*}
||\nabla f(x_1,x_2)||^2_2
&=&
(\partial_1 F(x)^2+\partial_2 F(x)^2)(\partial_3 F(x))^{-2} \\
&=&
\frac{4(1+r^2)}{r^2}\cdot\frac{\alpha^2 r^2(u(x_1,x_3)-a)^2+(1-\alpha(u(x_1,x_3)-a)^2)(1+r^2)}
{((1-\alpha(u(x_1,x_3)-a)^2)(1+r^2) +3 \alpha r^2(u(x_1,x_3)-a)u(x_1,x_3))^2} \\
&=&
\frac{4(1+r^2)}{r^2}\cdot\frac{\alpha^2 r^2(u(x_1,x_3)-a)^2+(1-\alpha(u(x_1,x_3)-a)^2)(1+r^2)}
{(\alpha(2r^2-1)(u(x_1,x_3)-a)^2+3\alpha a r^2(u(x_1,x_3)-a)+(1+r^2))^2} \\
&=&
\frac{4(1+r^2)}{r^2}\phi(t)
\end{eqnarray*}
where, for notational convenience, we have set
\[
\phi(t)= \frac{At^2+B}{(Ct^2+Dt+B)^2},
\]
with $t=u(x_1,x_3)-a$, $A=\alpha(\alpha r^2-1-r^2)$, $B=1+r^2$, $C=\alpha(2r^2-1)$, and $D=3\alpha ar^2$. Note that $B,C,D$ are all nonnegative while the sign of $A$ depends on $\alpha$. We show that $t\geq0$ implies
\begin{equation}
\label{eq: bound for phi 1}
\phi(t) \leq \phi(0) = \frac{1}{B} = \frac{1}{1+r^2}.
\end{equation}
To show this, assume first that $A\leq 0$. In this case, we easily see that $ABt^2+B^2\leq B^2\leq (Ct^2+Dt+B)^2$ so that~\eqref{eq: bound for phi 1} holds. Then, assume that $A>0$ and note that
\[
\phi'(t) = \frac{-2ACt^3+2B(A-2C)t-2BD}{(Ct^2+Dt+B)^3}.
\]
Since $\alpha<5$ and $r>1/\sqrt{5-\alpha}$, the numerator is strictly decreasing in $t$ and, hence, it is negative due to $-2BD\leq0$. Similarly, since $r>1/\sqrt{2}$, the denominator is strictly increasing in $t$ and, hence, it is strictly positive due to $B>0$. This establishes~\eqref{eq: bound for phi 1} also when $A>0$. As a result, we get
\[
||\nabla f(x_1,x_2)||^2_2 \leq \frac{4(1+r^2)}{r^2}\frac{1}{1+r^2} = \frac{4}{r^2}
\]
whenever $u(x_1,x_3)-a\geq0$, thus proving~\eqref{eq: bound gradient tilted 1}.

\smallskip

To prove~\eqref{eq: bound gradient tilted 2}, assume that $t$ belongs to the interval $[-1/\sqrt{\alpha},1/\sqrt{\alpha}]$. In this case, the numerator of $\phi(t)$ is easily seen to be maximized by $B$ if $A\leq 0$ and by $\tfrac{A}{\alpha}+B$ otherwise. At the same time, the denominator of $\phi(t)$ has its global minimum at $t=-\tfrac{D}{2C}$, which is larger or equal than $-1/\sqrt{\alpha}$ since $a^2\alpha\leq 1$ and $r>\sqrt{2}$. Hence, the denominator is minimized by $(B-D^2/4C)^2$. Now, if $A\leq 0$ we infer that
\begin{eqnarray*}
||\nabla f(x_1,x_2)||^2_2
&\leq&
\frac{4(1+r^2)}{r^2}\frac{16(1+r^2)(2r^2-1)^2}{((8-9\alpha a^2)r^4+4r^2-4)^2} \\
&=&
\frac{64(2r^4+r^2-1)^2}{r^2((8-9\alpha a^2)r^4+4r^2-4)^2} \\
&\leq&
\frac{256}{r^2(8-9\alpha a^2)^2}.
\end{eqnarray*}
The last inequality is due to the fact that, by assumption, $8-9\alpha a^2>0$ and $r\geq1$. On the other side, if $A>0$, then we have
\begin{eqnarray*}
||\nabla f(x_1,x_2)||^2_2
&\leq&
\frac{4(1+r^2)}{r^2}\frac{16\alpha r^2(2r^2-1)^2}{((8-9\alpha a^2)r^4+4r^2-4)^2} \\
&\leq&
\frac{64\alpha^2r^2(2r^2-1)^2}{((8-9\alpha a^2)r^4+4r^2-4)^2} \\
&\leq&
\frac{256\alpha^2r^2}{(1+r^2)^2(8-9\alpha a^2)^2}.
\end{eqnarray*}
The second inequality follows from $\alpha(1+r^2)\leq\alpha^2r^2$, which holds since $A>0$, and the last inequality is, as above, due to the fact that, by assumption, $8-9\alpha a^2>0$ and $r\geq1$. This finally establishes the bound in~\eqref{eq: bound gradient tilted 2}.

\subsubsection*{The basic set}

Consider the set $C\subset\R^2$ defined as the union of the following four quarters of ellipsoids:
\[
E_1 = \left\{(x_1,x_2)\in\R^2 \,; \ 4\left(x_1-\frac{1}{2}\right)^2+x_2^2\leq 1, \ x_1\in\left[\frac{1}{2},1\right], \ x_2\in[0,1]\right\},
\]
\[
E_2 = \left\{(x_1,x_2)\in\R^2 \,; \ \frac{4}{9}\left(x_1-\frac{1}{2}\right)^2+x_2^2\leq 1, \
x_1\in\left[-1,\frac{1}{2}\right], \ x_2\in[0,1]\right\},
\]
\[
E_3 = \left\{(x_1,x_2)\in\R^2 \,; \ 4\left(x_1+\frac{1}{2}\right)^2+x_2^2\leq 1, x_1\in\left[-1,-\frac{1}{2}\right], \ x_2\in[-1,0]\right\},
\]
\[
E_4 = \left\{(x_1,x_2)\in\R^2 \,; \ \frac{4}{9}\left(x_1+\frac{1}{2}\right)^2+x_2^2\leq 1, \
x_1\in\left[-\frac{1}{2},1\right], \ x_2\in[-1,0]\right\}.
\]

\begin{figure}[h]
\centering
\includegraphics[width=0.5\textwidth]{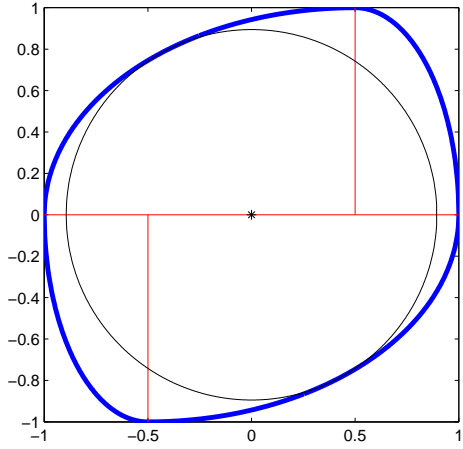}
\caption{The set $C$ as a union of four quarters of ellipsoids.}
\label{fig:asymmetric}
\end{figure}

Let $r>0$ be fixed and consider the set $\cB_r(C)$. This set has the same structure as the set $\cB_r$ considered above once we note that $\cB_r=\cB_r(E)$ for a suitable circle $E\subset\R^2$. The set $\cB_r(C)$ is clearly closed but its convexity is not a priori obvious.

\medskip

{\bf Convexity of $\cB_r(C)$}. We prove that $\cB_r(E_1\cup E_2)$ is convex. By symmetry, this will imply that $\cB_r(C)$ is also convex. Note that the half ellipsoid
\[
E'_1 = \left\{(u,v)\in\R^2 \,; \ 4\left(u-\frac{1}{2}\right)^2+v^2\leq 1, \ u\in[-1,1], \ v\in[0,1]\right\}
\]
is contained in the half ellipsoid
\[
E'_2 = \left\{(u,v)\in\R^2 \,; \ \frac{4}{9}\left(u-\frac{1}{2}\right)^2+v^2\leq 1, \ u\in[-1,1], \ v\in[0,1]\right\}
\]
so that $\cB_r(E'_1)\subset\cB_r(E_1\cup E_2)\subset\cB_r(E'_2)$. It follows from our previous paragraph that $\cB_r(E'_2)$ is convex. Now, take $x,y\in\cB_r(E_1\cup E_2)$ and note that any convex combination of $x$ and $y$ belongs to $\cB_r(E'_2)$. Since the equation
\[
u(\lambda x_1+(1-\lambda)y_1,\lambda x_3+(1-\lambda)y_3) = \frac{1}{2}
\]
has at most two solutions $\lambda\in[0,1]$, the segment with extremes $x$ and $y$ is divided in at most three subsegments where $u(x_1,x_3)-\frac{1}{2}$ has a constant sign. Positive subsegments are contained in $\cB_r(E'_1)$ and, hence, belong to $\cB_r(C)$. Negative subsegments are contained in \[
\cB_r(E'_2)\cap\left\{x\in\R^3 \,; \ -1\leq u(x_1,x_3)\leq\frac{1}{2}, \ x_2\geq 0\right\},
\]
which is also contained in $\cB_r(C)$. This establishes the convexity of $\cB_r(C)$.

\medskip

{\bf The boundary of $\cB_r(C)$ is smooth}. Let $F_1$ and $F_2$ be the functions associated to $\cB_r(E_1)$ and $\cB_r(E_2)$ as in~\eqref{eq: representation piece} and note that they coincide on the set $\{x\in\R^3 \,; \ u(x_1,x_3)=\tfrac{1}{2}\}$. Moreover, for every $x\in\R^3$ we have
\[
\nabla F_1(x)=
\left(\begin{array}{c}
4r^2x_3(2u(x_1,x_3)-1)\frac{\partial u(x_1,x_3)}{\partial x_1}\\
2x_2\\
4r^2(u(x_1,x_3)-1)u(x_1,x_3)+4r^2x_3(2u(x_1,x_3)-1)\frac{\partial u(x_1,x_3)}{\partial x_3}
\end{array}
\right)
\]
and
\[
\nabla F_2(x)=
\left(\begin{array}{c}
\frac{4}{9}r^2x_3(2u(x_1,x_3)-1)\frac{\partial u(x_1,x_3)}{\partial x_1}\\
2x_2\\
\frac{4}{9}r^2(u(x_1,x_3)-2)(u(x_1,x_3)+1)+\frac{4}{9}r^2x_3(2u(x_1,x_3)-1)\frac{\partial u(x_1,x_3)}{\partial x_3}
\end{array}
\right).
\]

\bigskip

Observe that $\nabla F_1(x)=\nabla F_2(x)$ whenever $u(x_1,x_3)=1/2$. Finally, take an arbitrary $(x_1,x_3)\in\R^2$ with $\tfrac{1}{2}\leq u(x_1,x_3)\leq1$ and $F_1(x_1,0,x_3)=0$. In this case, it is easy to see that $x_1=\sqrt{1+r^2x_3}$ so that, in fact, $u(x_1,x_3)=1$. In addition, we have $u(-x_1,x_3)=-1$ as well as
\[
\partial_1 u(-x_1,x_3)=\partial_1 u(x_1,x_3) \ \ \ \mbox{and} \ \ \ \partial_3 u(-x_1,x_3)=-\partial_3 u(x_1,x_3).
\]
As a result, we can use the above gradient formula to obtain
\[
\partial_1 F_2(-\sqrt{1+r^2x_3},0,x_3) = -\frac{12}{9}r^2x_3\partial_1 u(x_1,x_3)= -\frac{1}{3}\partial_1 F_1(\sqrt{1+r^2x_3},0,x_3),
\]
\[
\partial_2 F_2(-\sqrt{1+r^2x_3},0,x_3) = 0 = \partial_2 F_1(\sqrt{1+r^2x_3},0,x_3)
\]
\[
\partial_3 F_2(-\sqrt{1+r^2x_3},0,x_3) = \frac{12}{9}r^2x_3 \partial_3 u(x_1,x_3) = \frac{1}{3}\partial_3 F_1(\sqrt{1+r^2x_3},0,x_3).
\]
This shows that the boundary of $\cB_r(C)$ is smooth when $x_2=0$.

\medskip

{\bf ``Monotonicity'' of $\cB_r(C)$}. For any $0<R\leq\tfrac{7}{16}r$ consider the ice-cream cone
\[
\cK_R = \{x\in\R^3 \,; \ x_1^2+x_2^2\leq R^2x_3^2\}.
\]
We claim that
\begin{equation}
\label{eq: monotonicity twisted boat}
(\cB_r(C)+\cK_R)\cap\{x\in\R^3 \,; \ x_3\leq1\} = \cB_r(C).
\end{equation}

\smallskip

To show this, let $f_3$ be the function associated to $\cB_r(E_3)$ as in~\eqref{eq: implicit function} and note that $R\leq\tfrac{r}{2}$. As a result of the bound established in~\eqref{eq: bound gradient tilted 1}, it follows that
\[
R \leq \frac{1}{\|\nabla f_3(x_1,x_2)\|_2}
\]
for every $x\in\bd\cB_r(E_3)$ with $x_3>0$. The same bound holds, by symmetry, if we replace $E_3$ by $E_1$. Similarly, if $f_4$ is the function associated to $\cB_r(E_4)$ as in~\eqref{eq: implicit function}, then $R\leq\tfrac{7}{16}r$ implies that
\[
R \leq \frac{1}{\|\nabla f_4(x_1,x_2)\|_2}
\]
for every $x\in\bd\cB_r(E_4)$ with $x_3>0$ by~\eqref{eq: bound gradient tilted 2}. The same bound holds, by symmetry, if we replace $E_4$ by $E_2$. Then, one can easily establish~\eqref{eq: monotonicity twisted boat} following the lines of the proof of~\eqref{eq: monotonicity boat 0}.

\subsubsection*{The acceptance set}

Let $r=16$ and consider the convex acceptance set $\cA\subset\R^3$ defined by
\[
\cA = \Phi(\cB_r(C))+\R^3_+,
\]
where $\Phi$ is the rotation defined in~\eqref{eq: rotation}. Since $\cB_r(C)$ is a compact and convex set containing zero, we see that $\cA$ is closed, convex, and contains zero. In addition, $\cA$ satisfies by definition the monotonicity property. Hence, requirement (R1) is fulfilled.

\subsubsection*{The space of payoffs}

As above, the space of payoffs is the linear subspace of $\R^3$ given by
\[
\cM = \Phi(\cN)
\]
where $\cN=\{w\in\R^3 \,; \ w_2=0\}$. The pricing functional $\pi:\cM\to\R$ is defined by setting
\[
\pi(z) = z_3.
\]
In line with what established above, (R2) holds.

\medskip

Set $R=\sqrt{2}$ so that $\Phi^{-1}(\R^3_+)\subset\cK_R$. Since $R\leq\tfrac{7}{16}r$, it follows from~\eqref{eq: monotonicity twisted boat} that
\[
(\cB_r(C)+\Phi^{-1}(\R^3_+))\cap\{x\in\R^3 \,; \ x_3\leq1\} = \cB_r(C).
\]
Hence, we can reproduce the above argument to establish that $\risk$ is finite and (Lipschitz) continuous, so that (R3) holds as well. In other words, $(\cA,\cM,\pi)$ is an admissible triple.

\subsubsection*{The failure of continuous selections}

First of all, note that for every $x\in\R^3$ we can write
\begin{eqnarray*}
\Risk(x)
&=&
\{z\in\cM \,; \ z+x\in\cA, \ \pi(z)=\risk(x)\} \\
&=&
\{\Phi(w) \,; \ w\in\cN, \ w+\Phi^{-1}(x)\in\cB_r(C)+\Phi^{-1}(\R^3_+), \ \tfrac{w_3}{\sqrt{3}}=\risk(x)\} \\
&=&
\Phi(\{w\in\cN \,; \ w+\Phi^{-1}(x)\in\cB_r(C)+\Phi^{-1}(\R^3_+), \ \tfrac{w_3}{\sqrt{3}}=\risk(x)\}).
\end{eqnarray*}
Now, consider the sequence $(y^{(n)})\subset\R^3$ defined by
\[
y^{(2n-1)}=\left(0,-\frac{r}{\sqrt{n}},0\right) \ \ \ \mbox{and} \ \ \ y^{(2n)}=\left(0,\frac{r}{\sqrt{n}},0\right).
\]
Then, we easily see that
\[
\Risk(\Phi(y^{(2n-1)})) = \left\{\Phi\left(-\frac{1}{2}\sqrt{1+\frac{r^2}{n}},-\frac{r}{\sqrt{n}},\frac{1}{n}\right)\right\}
\]
and similarly
\[
\Risk(\Phi(y^{(2n)})) = \left\{\Phi\left(\frac{1}{2}\sqrt{1+\frac{r^2}{n}},\frac{r}{\sqrt{n}},\frac{1}{n}\right)\right\}
\]
for every $n\in\N$. Since $y^{(n)}\to0$ but we clearly have
\[
\left(-\frac{1}{2}\sqrt{1+\frac{r^2}{n}},-\frac{r}{\sqrt{n}},\frac{1}{n}\right)\to
\left(-\frac{1}{2},0,0\right) \ \ \ \mbox{and} \ \ \ \left(\frac{1}{2}\sqrt{1+\frac{r^2}{n}},\frac{r}{\sqrt{n}},\frac{1}{n}\right)\to
\left(\frac{1}{2},0,0\right),
\]
it follows that no selection of $\Risk$ can be continuous at $0$.


\end{document}